\documentclass[conference]{IEEEtran}
\IEEEoverridecommandlockouts
\usepackage{cite}
\usepackage{amsmath,amssymb,amsfonts}
\usepackage{algorithmic}
\usepackage{graphicx}
\usepackage{textcomp}
\usepackage{xcolor}
\usepackage{url}
\usepackage{stfloats}
\usepackage[caption=true,font=normalsize,labelfont=sf,textfont=sf]{subfig}
\usepackage{multirow}
\usepackage{hhline}
\usepackage{nicematrix,enumitem,booktabs}

\usepackage{titlesec}
\titlespacing*{\subsubsection}{0pt}{0.5ex}{0.25ex}
\titlespacing*{\subsection}{0pt}{0.5ex}{0.25ex}
\titlespacing*{\section}{0pt}{0.5ex}{0.25ex}
\usepackage{array}
\usepackage{ragged2e}
\newcolumntype{P}[1]{>{\RaggedRight\arraybackslash}p{#1}}

\begin{document}
\bstctlcite{IEEEexample:BSTcontrol}

\title{\fontsize{22pt}{24pt}\selectfont Lightweight Non-Line-of-Sight Channel Detection for ML-assisted Bluetooth Direction Finding
}

\author{\IEEEauthorblockN{Hamed Talebian\IEEEauthorrefmark{1}, Aamir Mahmood\IEEEauthorrefmark{1}, Mehdi Haghshenas\IEEEauthorrefmark{1}, Stefani Rydbloom\IEEEauthorrefmark{1}, Peter Karlsson\IEEEauthorrefmark{2}
Mikael Gidlund\IEEEauthorrefmark{1}}
\IEEEauthorblockA{\IEEEauthorrefmark{1}Department of Computer and Electrical Engineering,  Mid Sweden University, 851 70 Sundsvall, Sweden}
\IEEEauthorblockA{\IEEEauthorrefmark{2}Short Range Technologies, \textit{u-blox}, Malmö, Sweden}
\IEEEauthorrefmark{1}Email: \{firstname.lastname\}@miun.se,
\IEEEauthorrefmark{2}Email: peter.karlsson@ublox.com
\vspace{-15pt}
}

\maketitle

\begin{abstract}
Bluetooth Low Energy (BLE) direction-finding is promising for indoor industrial localization, but its accuracy degrades in multipath environments where reflections and scattering bias angle estimates. Although line-of-sight (LOS) and non-line-of-sight (NLOS) detection is well studied for wide-band radios, BLE direction-finding still lacks narrow-band channel-feature representations, scalable kernel-based feature transformations, and dedicated datasets for data-driven, lightweight channel classification. To address this gap, the work introduces a controlled BLE measurement setup that generates labeled LOS/NLOS data in two distinct propagation environments. A quality-driven machine learning (ML)-based pipeline is then developed for BLE Constant Tone Extension (CTE) In-phase-Quadrature (IQ) features. 
First, robust quantile-based standardization is applied to reduce the influence of outliers and heavy-tailed effects.
The standardized features are then analyzed using 
Principal Component Analysis (PCA) and Adaptive Kernel Density Estimation (AKDE) to verify scenario-dependent statistics and reveal LOS/NLOS separability.
Next, Nyström Kernel Approximation (NKA) constructs low-rank nonlinear feature maps followed by a lightweight Support Vector Classifier (SVC) head for LOS/NLOS detection. This classifier is compared with Random Forest (RF) and Multilayer Perceptron (MLP) models. Results show that NKA improves accuracy by about 7-14\% relative to the raw baseline. Although the MLP achieves higher absolute accuracy, the Nyström--SVC approach offers a more favorable trade-off between training complexity, inference cost, and memory footprint. Finally, several pipeline-calibrated posterior probabilities are utilized for cost-aware threshold selection and efficient real-time LOS/NLOS detection in resource-constrained localization systems.
\end{abstract}
\section{Introduction}
\IEEEPARstart{B}{LE} direction finding (DF) is widely used in industrial applications, including real-time tracking of tools, machinery, and personnel in smart factories, thereby improving inventory management and worker safety. In retail environments, BLE DF enables customer behavior analysis and location-aware services by accurately estimating shopper positions within stores \cite{u-blox2022ble}. Compared to vision- or infrared-based systems, radio-frequency sensing provides a ``versatile and non-intrusive” solution \cite{santra2025machine}. However, achieving high positioning accuracy and reliability generally requires a line-of-sight (LOS) path between the infrastructure and mobile units.

Increasing system bandwidth is shown to improve localization accuracy; for example, expanding bandwidth from 100\,MHz to 400\,MHz can reduce positioning error from the meter to the decimeter level under favorable conditions \cite[pp.~74--76]{project2025deliverable}. Nevertheless, such improvements do not fully mitigate errors in NLOS environments. Moreover, BLE angle-of-arrival (AoA), as a DF localization technique, operates within the constrained 2.4\,GHz ISM band, where bandwidth expansion is not feasible under current standards. Consequently, all wireless localization technologies, including BLE DF, are inherently susceptible to performance degradation in obstructed environments due to multipath effects.

In AoA-based localization, multipath-induced phase distortions directly bias inter-element phase measurements, leading to significant angle estimation error, necessitating robust mitigation strategies. To address this limitation, ML-assisted approaches have been increasingly explored to detect and subsequently discard or down-weight NLOS measurements, leveraging them as a quality indicator to enhance conventional triangulation algorithms. Prior work primarily focuses on ML for UWB and cellular systems, where physical-layer observables such as channel impulse response (CIR)-derived features (e.g., delay metrics and path power) enable effective channel classification \cite{bombino,olejniczak2024and,liu2022uwb}. However, these approaches are not directly applicable to BLE AoA estimation, where measurement modalities and hardware constraints differ fundamentally. In particular, there is a lack of rigorously validated ML frameworks tailored to BLE AoA measurements that operate on native observables such as CTE IQ statistics. Furthermore, publicly available BLE datasets (e.g., \cite{girolami2023experimental}) are not specifically designed for LOS/NLOS channel classification. Consequently, ML-assisted BLE localization requires feature representations and datasets adapted to the BLE channel characteristics.

Motivated by these limitations, this work proposes a lightweight ML-based pipeline with feature engineering tailored to BLE AoA estimation. To ensure both reliability and practical deployability, a comprehensive experimental framework is developed, including a controlled measurement campaign and a carefully designed setup to capture realistic LOS and NLOS propagation conditions. This enables the generation of representative, well-labeled measurements that reflect BLE AoA measurement characteristics. Furthermore, exploratory data analysis is conducted to reveal statistical separability between LOS and NLOS observations, thereby validating the feasibility of data-driven classification.

In practical deployments, BLE DF systems are typically implemented on resource-constrained embedded platforms, where critical constraints are computational complexity, memory footprint, and latency. Accordingly, the proposed framework emphasizes lightweight feature representations and low-complexity models to enable real-time operation without requiring specialized hardware acceleration. While this work focuses on BLE AoA measurements, the proposed methodology is general and can be extended to other short-range localization systems with similar limitations. Our proposal follows an \emph{identification-and-discard} approach, in which a computationally efficient LOS/NLOS detector identifies unreliable channels from CTE observations and excludes them from subsequent localization processing, while preserving reliable measurements for conventional angle estimation algorithms. The main contributions are:
\begin{itemize}
\item \textit{BLE LOS/NLOS dataset generation and statistical validation:} A controlled BLE AoA measurement campaign is conducted in two distinct propagation environments to generate labeled LOS/NLOS datasets. A PCA–AKDE-based statistical validation approach is developed to characterize class-conditional distributions for LOS/NLOS separability.
\item \textit{Kernel-feature selection:} The intrinsic dimensionality of the engineered feature space is analyzed, showing that Nyström kernel approximation (NKA) improves LOS/NLOS detection performance.
\item \textit{Quality-guided feature and model selection:} A systematic framework is proposed to select kernel types, hyperparameters, and feature budgets.
\item \textit{Low-complexity LOS/NLOS detection:} several lightweight classification pipelines are developed and compared to produce calibrated posterior probabilities, enabling thresholding under asymmetric cost of misclassifications for real-time resource-constrained device deployment.
\end{itemize}

\section{Methodology}
The main objective of the proposed framework is to jointly analyze LOS/NLOS channel characteristics and optimize ML-based detection performance.
Table~\ref{tab:methodology} summarizes the role of each mathematical tool in this workflow and the corresponding output used in the subsequent analysis.
\begin{table}[!ht]
\centering
\scriptsize
\caption{Mathematical tools for experimental data analysis}
\renewcommand{\arraystretch}{1.01}
\begin{tabular}{lll}
\hline
\textbf{Tool} & \textbf{Usage} & \textbf{output} \\
\hline
Learning curve & Compare training/inference time & Optimal training size  \\
FE & Overarching approach & Multiple pipelines \\
Scaler & Data scaling & Data normalization\\
PCA & Exploratory data analysis & Data verification \& KA \\
AKDE & Empirical PDF generation & Data verification \& QI \\
NKA & Non-linear learning & Feature space \\
Classifiers & NLOS detection & Label prediction \\
\hline
\end{tabular}
\label{tab:methodology}
\raggedright
{\emph{Note:} PDF: probability density func., KA: kernel approximation, QI: quality indicator}
\end{table}
\subsection{Learning curve}
\label{subsubsec:optimal_n}
Learning-curve analysis characterizes how generalization performance varies with training-set size \cite[pp.~708-711]{Perlich2017}. We use as the primary criterion the smallest sample size achieving at least $95\%$ of the maximum observed performance based on model accuracy.
\subsection{Feature Engineering (FE)}
\label{subsec:FE}
FE can be formalized as the design of a feature map \(T:\mathbb{R}^p\!\to\!\mathbb{R}^q\) applied row-wise to a dataset \(\mathbf{X}\in\mathbb{R}^{n\times p}\), producing \( \mathbf{U}=T(\mathbf{X})\in\mathbb{R}^{n\times q}\). In most practical pipelines, \(T\) is \emph{data-conditioned} through parameters, estimated from the training data and then held fixed for data transformation
Typical examples include standardization with \(\theta=(\mu,\sigma)\) and \(T_{\hat{\theta}}(x)=(x-\mu)/\sigma\). Such transformations are often composed as \(T=T_m\circ\cdots\circ T_1\) to improve conditioning, reduce variance, and align the data representation with the hypothesis space of the subsequent classification task \cite{HastieESL}.
A unifying view is that FE increases the expressiveness of simple predictors by constructing representations where a linear decision function approximates a nonlinear structure. This highlights the importance of FE, as it enables data representations that favor simple, preferably linear, and lightweight classifiers, making them suitable for deployment on resource-constrained devices.

\subsubsection{Principal Component Analysis (PCA)}
\label{subsub:pca}
PCA is particularly advantageous for exploratory label verification, as inherent mutual coupling in the BLE receiver antenna array introduces significant spatial correlation. It serves as a robust tool to transform a set of correlated observations into a series of uncorrelated principal components (PC) through an orthogonal transformation.  Let the centered LOS and NLOS feature vectors be denoted by $\mathbf{x}_n^{(L)}\in\mathbb{R}^p$ and $\mathbf{x}_n^{(NL)}\in\mathbb{R}^p$, respectively. PCA first computes the sample mean (zero mean dataset generation) and approximates the covariance matrix from the training data, then extracts the $r$ dominant eigenvectors 
(See. \cite[ch.~19]{BishopPRML}). Each sample is then projected onto the principal subspace by taking inner products with these axes. 
Stacking the projected score vectors row-wise yields the class-specific PCA matrices ($\mathbf{Z}_{\mathrm{L/NL}}$) 
which are the low-dimensional representations of the original observations.

\subsubsection{Adaptive Kernel Density Estimation (AKDE)}
\label{subsub:AKDE}
AKDE is employed, when required, to construct an effective empirical PDF that provides a smooth and data-adaptive approximation of the underlying distribution inferred from histogram-based representations. Abramson’s AKDE \cite{abramson1982} extends the fixed-bandwidth Gaussian KDE by assigning a local bandwidth to each sample, which reduces over-smoothing in dense regions and stabilizes estimation in sparse tails. 
This yields narrower kernels in high-density and wider ones in low-density regions, providing improved modeling of skewed or heavy-tailed data, which is expected to be observed in NLOS.

\subsubsection{Nyström Kernel Approximation (NKA)}
NKA is employed in this paper to construct computationally efficient feature mappings enabling lightweight pipelines suitable for resource-constrained BLE devices, while effectively capturing the nonlinear structure of the underlying dataset. 
KA enables nonlinear learning by mapping input data into a high-dimensional space via a kernel, where computations rely on the Gram matrix. Common kernels are presented in Table~\ref{tab:kernel_choices}. 
Quadratic storage and computational cost 
limits KA scalability. The Nyström method is employed to mitigate this issue by constructing a low-rank approximation of the full kernel matrix. NKA selects a small subset of $r<<n$ representative data points, referred as landmark points, to evaluate pairwise similarities across all $n$ samples, then, approximates the global kernel structure by computing similarities between the full dataset and these landmarks, as well as among the landmarks themselves. 
The approach significantly reduces complexity from $\mathcal{O}(n^2)$ to $\mathcal{O}(nr)$ and its quality is evaluated using metrics such as centered kernel alignment (CKA) and relative Frobenius error \cite{cristianini2001alignment,cortes2012cka}.
\begin{table}[!ht]
\caption{Kernel choices for Nyström KA technique}
\label{tab:kernel_choices}
\centering
\scriptsize
\renewcommand{\arraystretch}{1.30}
\begin{tabular}{p{0.45\columnwidth} p{0.45\columnwidth}}
\hline
\textbf{Item} & \textbf{Equation} \\\hline
Linear &
$K_{\text{lin}}(x,y)=x^{\top}y$\\
\renewcommand{\arraystretch}{1}
Radial Basis Function (RBF) &
$K_{\text{rbf}}(x,y)=\exp\!\big(-\gamma\|x-y\|_2^2\big)$\\
Random Fourier Features (RFF) &
Randomized RBF KA \\
Polynomial (degree 3) &
$K_{\text{poly}}(x,y)=\big(\gamma\,x^\top y + c_0\big)^3
$\\
Sigmoid (tanh) &
$K_{\text{sig}}(x,y)=\tanh\!\big(\gamma\,x^\top y + c_0\big)
$\\\hline
\end{tabular}
\label{tab:kernel_choice}
\end{table}
\subsection{Lightweight Classifiers} Since BLE DF systems are commonly deployed on resource-constrained platforms, LOS/NLOS detection should be performed with low inference latency, limited memory footprint, and minimal additional processing. For this reason, the proposed framework considers compact classifiers that operate on engineered or kernel-transformed CTE-IQ features, and their performance is compared with two baselines, including MLP and RF. MLP provides strong nonlinear modeling capability by learning hierarchical feature representations directly from data (see  \cite{Shrestha2025DLReceiverSurvey}); however, this comes at the cost of significantly higher training complexity and sensitivity to hyperparameter selection. In contrast, SVC (see \cite{griva2008linear}) offers a convex optimization solution, but its linear formulation yields inadequate accuracy, requiring NKA to enhance performance, while offering a favorable balance between accuracy and complexity. RF (see \cite{breiman2001random}) demonstrates fast training and low inference complexity, making it particularly suitable for real-time applications, while it is not memory efficient given its rule-based ensemble model. The aforementioned advantages and limitations are corroborated by the theoretical analysis in Table~\ref{tab:complexity_comparison} and the empirical results in Fig.~\ref{fig:perf_comp}c and Fig.~\ref{fig:perf_comp}d.

\section{Experimental setup}
\label{sec:experimental_setup}
This section presents the experimental scenarios measurement campaign, and dataset characteristics.

\subsection{Hardware}
All data were collected using the u-blox XPLR-AOA-3 development kit 
seen in Fig.~\ref{fig:kit}. It consists of ANT-B10 anchor nodes equipped with an eight-element planar patch antenna array and an integrated NINA-B411 BLE 5.1 direction-finding radio, together making up the anchor point (AP), and a C209 BLE transmitter (tag). During all measurements, the transmit power of the tag was fixed at $1$\;mW. For each measurement position, 1000 CTE packets were recorded. Since the receiver exposes IQ samples associated with each CTE, all subsequent analysis is performed directly on the IQ-derived observations.

NLOS conditions were enforced by inserting a graphene-coated blocker on a wooden stand, as seen in Fig.~\ref{fig:block}, between the tag and the active anchor. It was dimensioned and designed to eliminate the direct LOS component of the signal from the tag, while allowing only reflected/scattered components from the environment to reach the anchor.

\subsection{Experimental Scenarios}
Measurements were conducted in two distinct indoor environments in the same building on the Mid Sweden University campus, to form two different scenarios.

\begin{figure}[t]
\centering
\subfloat[]{\fbox{\includegraphics[width=0.39\columnwidth]{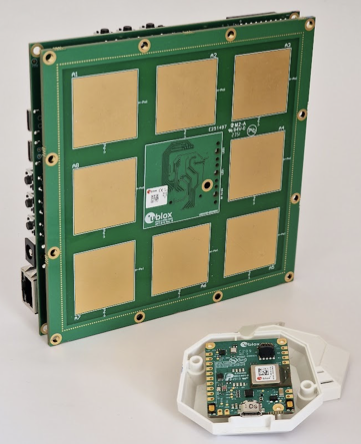}}\label{fig:kit}}
\hfill
\subfloat[]{\fbox{\includegraphics[width=0.40\columnwidth]{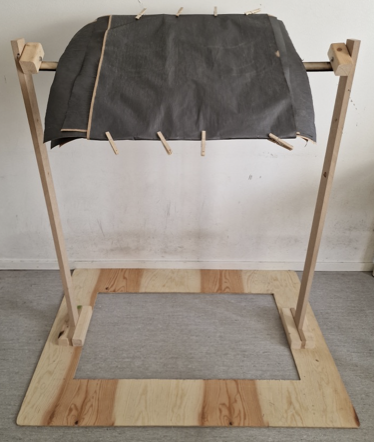}}\label{fig:block}}
\caption{Development kit by u-blox (a), and graphene signal blocker (b).}
\label{fig:exp_setup}
\subfloat[]{\fbox{\includegraphics[width=0.40\columnwidth]{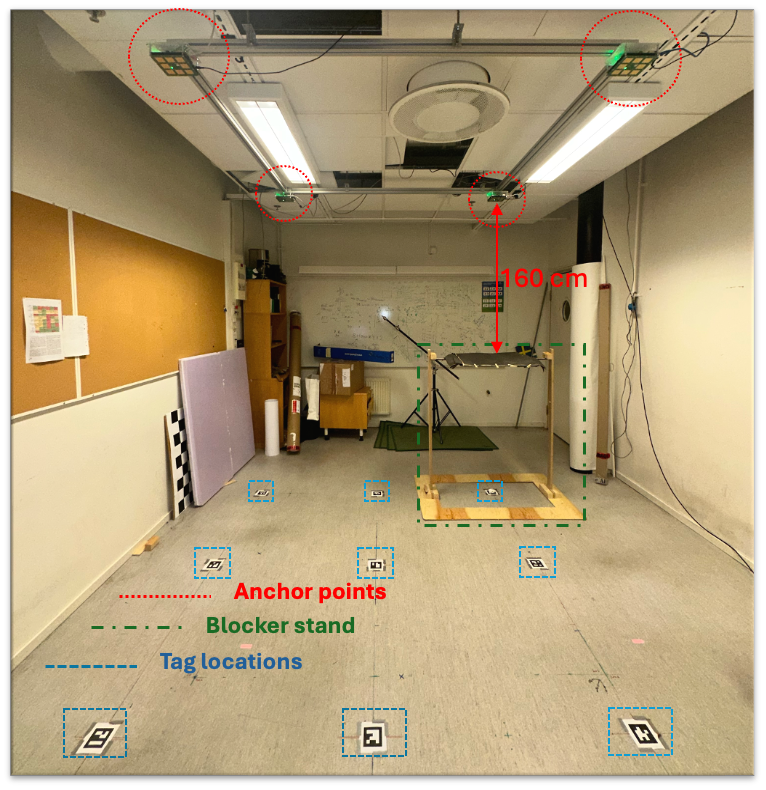}}\label{fig:R}}
\hfill
\subfloat[]{\fbox{\includegraphics[width=0.40\columnwidth]{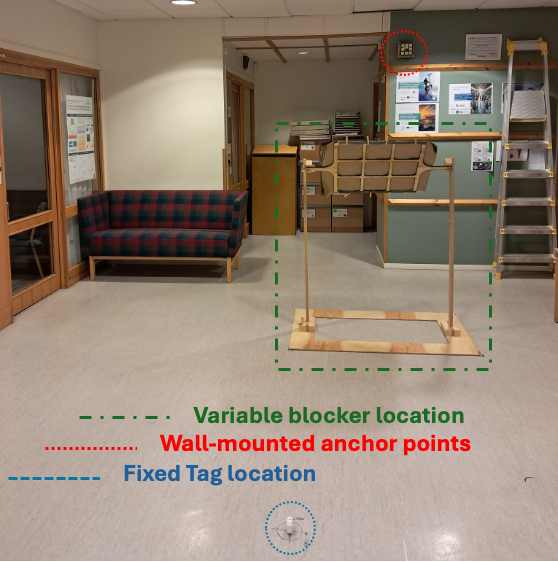}}\label{fig:O}}
\caption{Measurement environment in Scenario ``\textit{R}” (a), and ``\textit{O}” (b).}
\label{fig:env}
\end{figure}

Scenorio ``\textit{R}", was conducted in a square room, seen in Fig.~\ref{fig:R}, with horizontal dimensions $630 \times 385$\,cm. It has epoxy-covered concrete flooring and normal building material in the walls, ceiling, and furniture containing metal parts as well as a cylindrical metal pillar forming at part of the building structure. No attempt was made to dampen the reflective effect of these metal features, as they are likely to exist in any indoor space. 
In ``\textit{R}", data were collected using four antenna arrays, mounted in four positions on an aluminum frame, all facing downward. The transmitting tag was placed in five positions that form an L-pattern directly below each receiving antenna, and the blocker was positioned perpendicular to the LOS of the receiving anchor points during the NLOS loggings.

The space in scenario ``\textit{O}", seen in Fig.~\ref{fig:O}, was a slightly larger ($947 \times 464$\,cm) space between office rooms, with three APs placed on different walls facing the center of the room. In ``\textit{O}", the transmitting tag was placed in a single position in the center of the room. The blocker was positioned in a similar way as in scenario ``\textit{R}", with the blocking surface perpendicular to the LOS of the receiving anchor points.

The amount of data labeled ``LOS” is equal to ``NLOS”. However, since the number of tag positions and the number of anchor points were higher in scenario ``\textit{R}", the amount of data in scenario ``\textit{R}" is larger than in scenario ``\textit{O}".

\subsection{Preprocessing and Evaluation Protocol}
\label{evaluation_protocol}
Each recorded CTE packet is represented through an IQ-derived feature vector and processed independently. To broaden the evaluation beyond the individual ``\textit{R}” and ``\textit{O}” datasets, we additionally construct two synthetic combined datasets, denoted as ``\emph{Mixed (M)}” and ``\emph{Proportional (P)}”. The ``\emph{M}” dataset is obtained by concatenating ``\emph{R}” and ``\emph{O}” CTE-packet collections and then randomly selecting 54k packets from the resulting pool. This setting is used to examine classifier behavior under pooled multi-environment observations and to assess whether increased sample heterogeneity improves robustness. The ``\emph{P}” dataset is designed to reduce environment imbalance. Because the ``\emph{R}” dataset size is larger than ``\emph{O}”, we randomly down-sample ``\emph{R}” to the size of ``\emph{O}” and concatenate the two equally sized subsets. This construction yields a balanced dataset with comparable contributions from both propagation conditions, enabling a fairer assessment of generalization across the measured environments.
\subsection{Dataset characteristics}\label{subsec:data_character}
The original experimentally collected datasets consist of $\mathbb{R}^{132\text{K}\times 64}$ for the ``\emph{R}” scenario and $\mathbb{R}^{54\text{K}\times 64}$ for the ``\emph{O}” scenario. The general dataset statistics are presented in Table.~\ref{tab:dataset_stats}. In all cases, each row corresponds to a single CTE packet acquired during the measurement campaign and contains 64 absolute IQ-derived values determined by the eight-element antenna array and the employed switching pattern of the BLE device. Each observation is labeled as (-1) or (+1), corresponding to LOS and NLOS, respectively. After constructing the ``\emph{M}” and ``\emph{P}” datasets, each is partitioned such that 70\% is used for training and the remaining is reserved for testing, with benchmark subsets of size 3K and 7K from training sets (see \ref{subsubsec:optimal_n}). 
\begin{table}[!ht]
\centering
\caption{Dataset general statistics}
\label{tab:dataset_stats}
\begin{tabular}{lll} \hline 
\scriptsize
\renewcommand{\arraystretch}{0.7}
 \textbf{Item} &  \textbf{\textit{``R”}} &  \textbf{\textit{``O”}}  \\
 \hline
  Transmitter positions & 12 & 1 \\
  Blocker positions & 12 & 3 \\
Samples per event & 1000  & 1000 \\  
  packets per AP & 54\;K & 18\;K \\ 
 data packets & 216\;K & 54\;K \\ 
  IQ values & 15,552\,K & 3,888\,K  \\ 
        LOS samples & 108\,K & 24300  \\
               NLOS samples & 108\,K & 24300\\
        Training samples & 97200 & 48600 \\ 
    Testing samples & 108200 & 24300 \\
               CTE length & 64 & 64\\\hline
\end{tabular}
\end{table}
Whenever validation or cross-validation is required, the validation subset is drawn randomly from the training partition. Moreover, all exploratory data analysis is performed using the whole training data, while random subsets of the training set are used for hyperparameter optimization and AKDE-based analysis when needed. 

\section{Results}\label{sec:results}
The pipeline configuration and the multi-performance comparison among 432 training tasks are presented in this section.

\subsection{Baseline Block: Scaling and Diagnostics}
\label{subsec:scaling_block}

Column-wise normalization operators are fitted on all datasets, including \textit{StandardScaler (S)} and \textit{RobustScaler (Ro)} in Scikit-learn libraries.
The robust variant is parameterized by the Inter quantile Range (IQR) of 25\% and 75\%, and both methods are evaluated with a common outlier threshold ($\tau=1.28$). 
To select the first-stage block on the pipeline, two sets of diagnostics are computed on the scaled datasets. Both operators achieve essentially perfect centering and normalization across all datasets, with near-zero centering error and unit scale variance in all cases. The main difference is that Ro reduces the observation size to about 0.01 across all datasets, while maintaining a canonical $\mathrm{IQR}=0.35$, whereas S outlier fraction is about 0.13-0.14. Given an insignificant difference, Ro is selected as the first FE block. 
\subsection{Exploratory Data analysis}
\label{subsec:exploratory_data_analysis}
To compare the statistical structure of the IQ-power data in the ``\emph{R}” and ``\emph{O}” scenarios, each dataset is independently standardized and projected into an 8-dimensional PC space, explaining 85\% of the variance for both datasets. 
Normalized cross-correlation analysis between ``\emph{R}” and ``\emph{O}” PC representations yields values on the order of $10^{-15}$, indicating negligible linear dependence. This confirms that the two scenarios are statistically independent, with distinct variability structures in the transformed feature space.

\begin{figure}[!t]
    \centering
    \includegraphics[width=1\columnwidth]{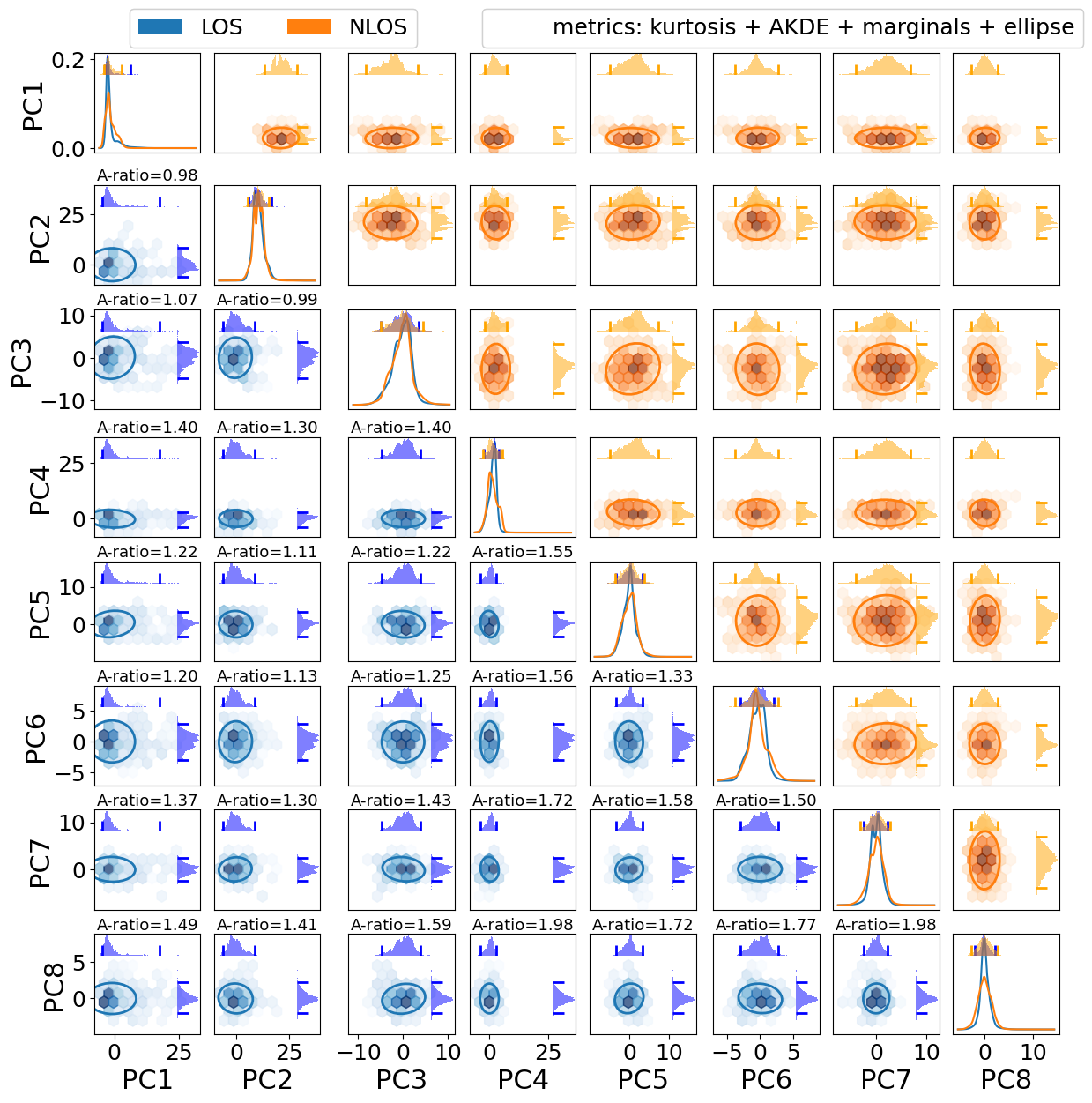}
    \caption{Class-conditional pairwise PCA diagnostics of CTE features for ``\emph{O}” scenario. Off-diagonal hexbin plots visualize 2D data density by grouping points into hexagonal bins and encoding their kurtosis via color. $\mathrm{grid\_size}=10$ controls the resolution of hexagonal bins, while $\mathrm{mincnt}=20$ filters out low-density bins. To capture a fraction of the distribution, a confidence ellipse is generated by computing the sample mean and covariance of the 2D data and drawing the contour satisfying $(x-\mu)^\top\Sigma^{-1}(x-\mu)=\chi^2(0.95)$, where $\chi^2$ is the chi-square distribution. Marginals are the individual one-dimensional distributions of each principal components of LOS and NLOS obtained by integrating the joint distribution over the other variables.}
    \label{fig:kde_comparison}
    \vspace{-2pt}
\end{figure}

To characterize LOS/NLOS differences after feature orthogonalization, the preprocessed IQ-power vectors are projected onto PC space and analyzed using pairwise diagnostics. Fig.~\ref{fig:kde_comparison} depicts LOS and NLOS distributions on principal feature space for ``\emph{O}” scenario. The diagonal panels show AKDE distributions and class-conditional marginal densities (See. \ref{subsub:AKDE}), while off-diagonal panels use hexagonal binning colored by local excess kurtosis, $\eta=\mu_4/\mu_2^2-3$, where $\mu_2$ and $\mu_4$ are second and forth moments. From the diagonal panels, the shapes of the AKDE empirical PDFs differ between LOS and NLOS, with NLOS exhibiting heavier tails that become more evident as the number of PC increases; as expected, incorporating additional PC captures greater variance, thereby enhancing their separability. Similarly, the separability between LOS and NLOS becomes more pronounced as the number of PCs increases, while the dispersion of dense hexagonal regions increases for NLOS in the off-diagonal panels. This phenomenon is quantified by the $\mathrm{A-ratio} = A_{\mathrm{NLOS}}/A_{\mathrm{LOS}}$, where $A_{(.)}$ denotes the respective ellipse areas. As seen, this value is above one for higher-order PCs, which is relevant since NLOS measurements often exhibit heavier tails and outlier-prone behavior due to multipaths. The ``\emph{R}” scenario has the same general trend, but indicating less separability than ``\emph{O}” scenario, whereas pairwise projections reveal broader and less regular NLOS support, especially in higher-order PCs.

\subsection{Pipeline Configuration}
\label{subsec:pipeline_configuration}
Our objective is to configure all pipelines such that they utilize the minimum number of trainable parameters while preserving acceptable classification performance. To this end, a baseline MLP with two hidden layers of sizes 16 and 8 is first considered, achieving approximately 90\% accuracy averagely, thereby establishing a reference for lightweight nonlinear modeling. For SVC-based pipelines, hyperparameter tuning is performed via grid search, where the feature budget is constrained between 32 and 64 components. The results indicate that NKA approaches unity at 64 components, suggesting sufficient representational capacity within this range. Across all SVC-based configurations, the selected kernel bandwidth remains consistently close to $6\times10^{-4}$. Additionally, an eight-dimensional linear PCA kernel is evaluated as a lower-bound baseline. For the RF baseline, the number of trees is fixed to 100, with eight candidate features per split and a maximum depth of five, yielding a model with comparable complexity to the other pipelines. All pipeline configurations maintain parameter counts within the same order of magnitude, supporting lightweight deployment, as summarized in Table~\ref{tab:complexity_comparison}.

\begin{table}[!ht]
\centering
\scriptsize
\caption{Pipeline configuration comparison}
\label{tab:complexity_comparison}
\renewcommand{\arraystretch}{0.5}
\begin{NiceTabular}{lcccc}
\hline
\textbf{Model} & \textbf{Parameters} & \textbf{Tr. FLOPs}\tabularnote{Training FLOPs is computed for 3k sample size and 2k iteration} & \textbf{Inf. FLOPs}\tabularnote{Inference FLOPs is computed for a single sample} & \textbf{Memory (KB)}\tabularnote{Memory calculation is based on float32=4\;byte} \\
\hline
MLP 
& $1.2 \times 10^3$ 
& $2.1 \times 10^{10}$ 
& $1.2 \times 10^3$ 
& $4.80$ \\
SVC 
& 65 
& $3.8 \times 10^8$  
& 64
& 0.26 \\
RFF\tabularnote{RFF parameters are fixed and stored once (a single forward pass without iteration), while each sample/input must mapped to the RFF space before SVC operation. Doubling the features recovers the full RBF representation.}  ($n$=64)
& 0 
& $1.2 \times 10^7$ 
& $4.1 \times 10^3$ 
& 17.00 \\
RFF + SVC
& $ 65$ 
& $4.0 \times 10^8$ 
& $4.2 \times 10^3$ 
& 17.26 \\
RF\tabularnote{Since RF is a rule-based (non-parametric) model, the notions of parameters and FLOPs are not strictly applicable; however, approximate counts are used for consistent comparison with other classifiers. For RF memory estimation, it is assumed that each node stores five rule-related values.} 
& $5-6\times10^3$ 
& $1.2 \times 10^7$ 
& $5.0 \times 10^2$ 
& $100$--$120$KB \\
\hline
\end{NiceTabular}
\end{table}
\subsection{Classification Performance}
\begin{figure}[!ht]
    \centering
    \subfloat[\tiny Accuracy]{
        \includegraphics[width=1\columnwidth]{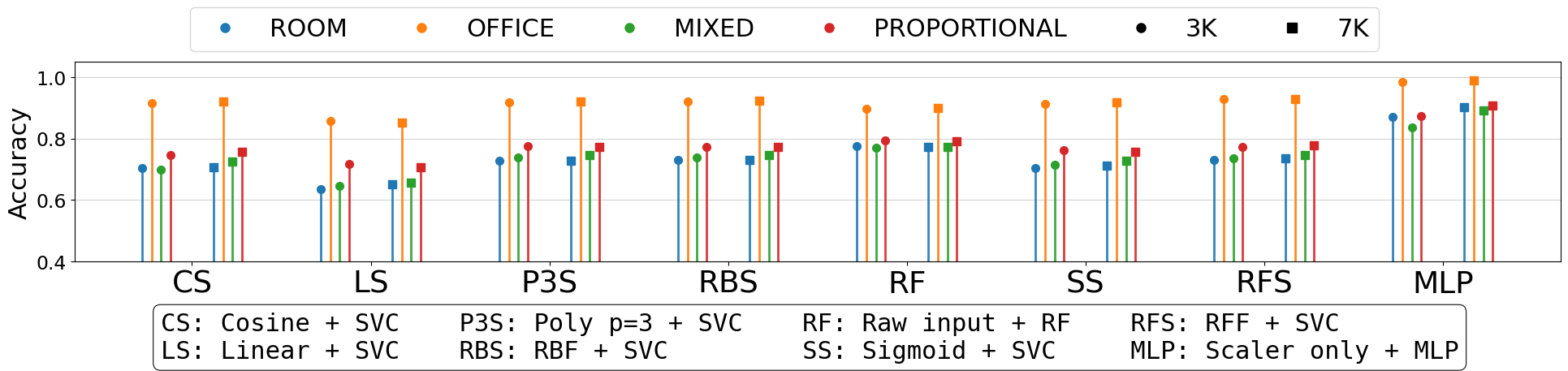}
        \label{bar_acc}
    } 
    \hspace{0.0\textwidth} 
    \subfloat[\tiny Accuracy with synthetic AWGN]{
        \includegraphics[width=1\columnwidth]{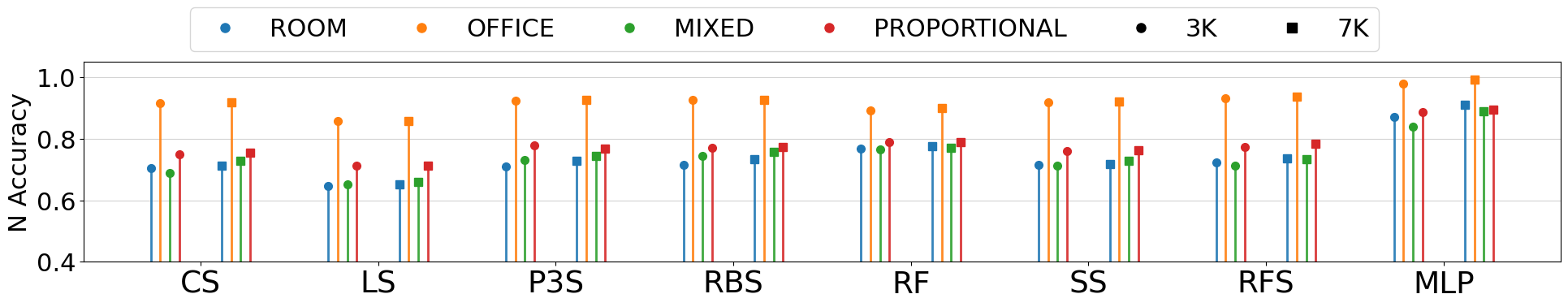}
        \label{bar_kappa}
    }
    \hspace{0.0\textwidth}
    \subfloat[\tiny Training time]{
        \includegraphics[width=1\columnwidth]{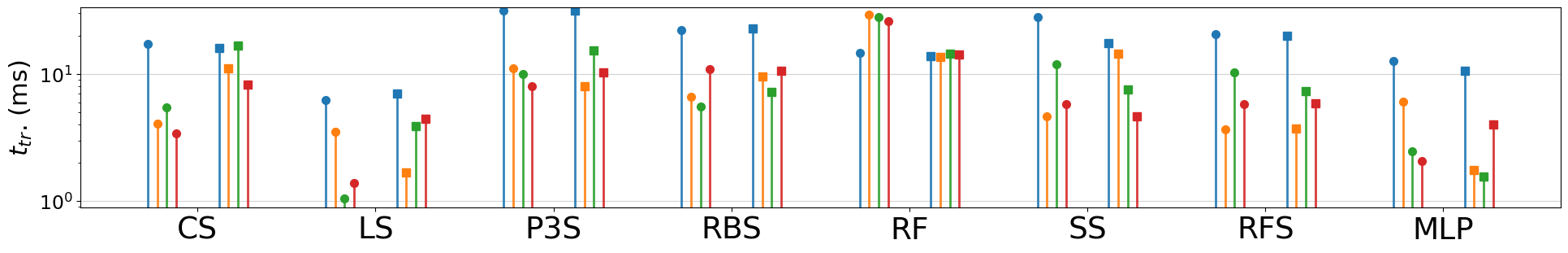}
        \label{bar_fit_time}
    }
    \hspace{0.0\textwidth}
    \subfloat[\tiny Prediction time]{
        \includegraphics[width=1\columnwidth]{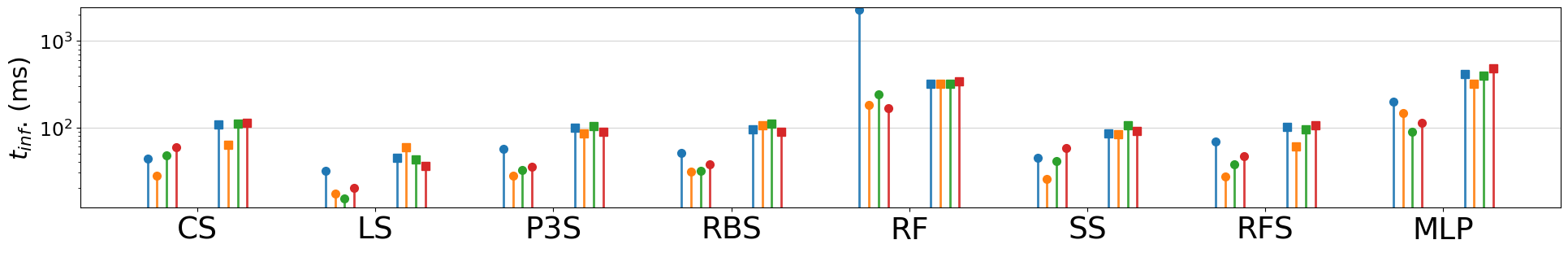}
        \label{fig:3}
    }
    \caption{Multi-criteria pipeline comparison for four datasets and 3K/7K subsets. \textcolor{black}{In (b), the model's robustness to noise is evaluated by synthetically adding additive AWGN with SNR = 20\;dB to the original datasets and measuring the resulting accuracy. The observed performance degradation is limited to approximately 1–3\%, indicating that the models exhibit insensitivity and are not overfitting to noise perturbations.}}
    \label{fig:perf_comp}
\end{figure}

Classification performance is summarized for 3K and 7K training size across four datasets, described in section~\ref{subsec:data_character}. Fig.~\ref{fig:perf_comp}a represents accuracy values, which show a clear ranking. The near-perfect performance observed in the \textit{O} scenario, with accuracy approaching 98\% to 99\% for the 3k and 7k datasets, reflecting both the strong discriminative capability of MLP and the comparatively simpler, more structured propagation environment. The \textit{R} dataset follows, improving from  87\% at 3K to 91\% at 7K. The two combined datasets are more challenging, but still strong: \textit{M} reaches 84\%–89\%, while \textit{P} reaches 87\%–91\%. Thus, among the pooled datasets, the \textit{P} has consistently and marginally higher performance than the \textit{M} dataset, which supports the claim that more balanced and diverse training composition improves prediction generalization, although it reduces the overall performance compared with \textit{R} and \textit{O} datasets. 
NKA consistently improves accuracy relative to the linear-SVC baselines. For example, at 7K, linear SVC gives $\mathrm{Acc.}= \{65\%, 86\%, 66\%, 71\%\} \in \{\mathrm{\textit{R, O, M, P}}\}$, whereas the best NKA-based SVC variants increase these values to about $\mathrm{Acc.}=\{79\%, 93\%, 73\%, 78\%\}$, indicating 7\% to 14\% increase in discriminative performance. 

Across all datasets and evaluation scenarios, scaled MLP, RF, and RFF+SVC pipelines consistently achieve the highest accuracy, although their relative ranking varies depending on the specific conditions.
\textcolor{black}{Table.~\ref{tab:desgin_perspective} presents the best pipelines in terms of accuracy and training/inference time ($t_{tr}/t_{inf}$), supporting the use of FE for BLE LOS/NLOS channel detection. As seen, the difference in accuracy score is marginal for RF and RFF+SVC, while SVC reduces $t_{inf}$ by approximately 44\%, identical to that of MLP. Moreover, RFF+SVC accuracy is 7\% to 11\% higher than linear and direct SVC implementation. Even when the classification accuracy of the SVC branch reaches a moderate-to-high regime, the NKA architecture remains appealing, as it enhances accuracy without introducing significant implementation complexity while maintaining low $t_{inf}$ and substantially reduced $t_{tr}$. In other words, explicit kernel maps such as NKA and especially RFF provide a practical, balanced trade-off between accuracy, memory, and computation. This makes them suitable for resource-constrained deployment, where a moderate but reliable increase in NLOS detection is more desirable than a marginally higher accuracy at a significantly greater computational cost.}
\begin{table}[!t]
\centering
\scriptsize
\caption{Multi-criteria pipeline performance comparison}
\label{tab:pipeline_comparison}
\scriptsize
\renewcommand{\arraystretch}{1}
\begin{tabular}{lrrrrr}
\toprule
\textbf{Model} & \textbf{Accuracy} & \textbf{TPR} & \textbf{TNR} & \textbf{$\mathbf{t_{tr.}}$\;(s)} & \textbf{$\mathbf{t_{inf.}}$\;(s)} \\
\midrule
Scaler only + MLP & 0.908 & 0.908 & 0.907 & 0.602 & 0.009 \\
Raw input + RF & 0.809 & 0.840 & 0.778 & 0.269 & 0.016 \\
RFF + SVC & 0.792 & 0.820 & 0.764 & 0.053 & 0.009 \\
RBF + SVC & 0.791 & 0.798 & 0.783 & 0.057 & 0.016 \\
Poly p=3 + SVC & 0.786 & 0.791 & 0.782 & 0.060 & 0.017 \\
Sigmoid + SVC & 0.779 & 0.778 & 0.780 & 0.059 & 0.013 \\
Cosine + SVC & 0.776 & 0.765 & 0.787 & 0.061 & 0.009 \\
Raw input + MLP & 0.763 & 0.748 & 0.778 & 0.358 & 0.002 \\
Linear + SVC & 0.720 & 0.749 & 0.690 & 0.041 & 0.006 \\
Raw input + SVC & 0.685 & 0.686 & 0.684 & 0.075 & 0.003 \\
\bottomrule
\end{tabular}
\raggedright
{\emph{Note:}\;values are averaged over all dataset types and dataset sizes. \\
\emph{Note:}\;TPR: True Positive Rate; TNR: True Negative Rate}
\label{tab:desgin_perspective}
\end{table}

\section{Discussion}
\begin{figure}[!ht]
    \centering
\includegraphics[width=0.8\linewidth]{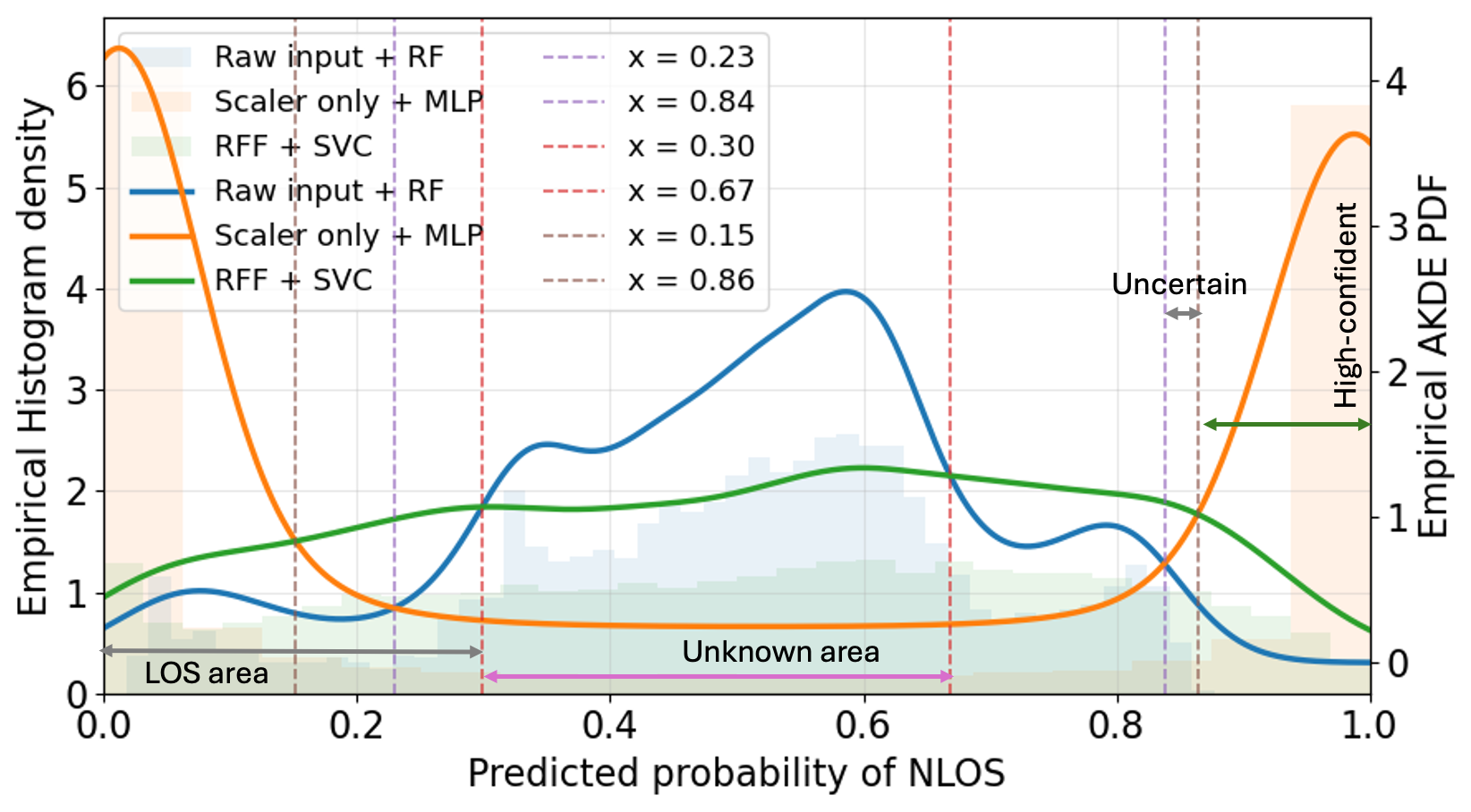}
    \caption{\textcolor{black}{Normalized probability density function (PDF) is generated by computing the 
Adaptive Kernel Density Estimation as $\hat{f}(x) = n^{-1}\sum_{i=1}^{n} \varphi((x-X_i)h_i^{-1})$, where $X_i$ is observed data points and $x$ is a query point. Then it is normalized by
$f_{\text{norm.}}(x)=\hat{f}(x)/\int \hat{f}(x)\,dx$ to ensure unit area under the curve and valid PDF properties. 
}}
    \label{fig:AKDE_intersections}
\end{figure}

The current results highlight broader implications for ML-based localization:
\begin{itemize}
    \item Probabilistic prediction analysis can be employed as a quality indicator and as a basis for adaptive threshold setting for an \textit{identification-and-discard} strategy. For instance, AKDE intersections and local peak structure provide data-driven operating points at which the decision threshold can be adjusted according to the desired trade-off among sensitivity, specificity, and cross-pipeline agreement through a majority voting mechanism. An example is presented in Fig.~\ref{fig:AKDE_intersections}. Empirical parametric functions are obtained from the predicted NLOS probabilities of the selected pipelines, whose location-dependent bandwidth is based on AKDE. 
    The pairwise intersections of these curves mark thresholds at which two pipelines assign equal density, thereby serving as compact indicators of score overlap, quality, and calibration consistency.
    \item Incorporate link-quality estimation into the BLE positioning engine to assign confidence-aware prediction weights to AoA measurements, enabling weight-dependent triangulation and improved robustness.
\end{itemize}

\section{Conclusion}
In this work, a lightweight ML-assisted BLE direction-finding framework for LOS/NLOS identification was developed and evaluated under four tightly controlled datasets constructed from BLE CTE IQ measurements. The study showed that robust preprocessing and feature engineering are essential for BLE-native LOS/NLOS discrimination, with PCA-based diagnostics confirming scenario-dependent statistics and stronger class separability in the ``\textit{Office}” environment than in the ``\textit{Room}” case. Quality-guided Nyström Kernel approximation, together with compact kernel feature maps and calibrated classifiers, consistently improved LOS/NLOS detection over linear baselines, while RFF emerged as the most reliable KA across datasets. Although RF and MLP achieved the highest absolute performance, the Kernel-based linear SVC provided a favorable balance between agreement, runtime, and memory, making it attractive for resource-constrained \textit{identify-and-discard} deployment, with moderate to substantial class separability. Overall, the results confirm that BLE-specific feature transformation is a practical and effective path toward robust NLOS \textit    {classification-and-discard} in industrial BLE AoA localization, while also motivating future extensions toward adaptive thresholding and confidence-aware LOS/NLOS function for a localization engine.

\bibliographystyle{IEEEtran}  
\bibliography{references}

@IEEEtranBSTCTL{IEEEexample:BSTcontrol,
  CTLuse_forced_etal       = "yes",
 CTLmax_names_forced_etal = "2",
  CTLnames_show_etal       = "1" 
}

@techreport{u-blox2022ble,
  author    = {Peter Karlsson},
  title     = {Getting started with Bluetooth for high precision indoor positioning},
  year      = {2021},
  institution = {u-blox AG},
}

@techreport{project2025deliverable,
  author       = {{5G Smart Manufacturing}},
  title        = {Deliverable {D5.3}: Second Report on new Thechnological Features to be supported by {5G} standardization and their implementation impact},
  institution  = {EU Horizon 2020},
  number       = {D20},
  year         = {2021},
  month        = {Nov},
}

@article{santra2025machine,
  title={Machine Learning-Powered Radio Frequency Sensing: A Review},
  author={Santra, Avik and Wang, Pu and Shaker, George and Mysore, Bhavani Shankar and Dolmans, Guido and Chen, Yan and Shariati, Negin and Pandharipande, Ashish},
  journal={IEEE Sensors J.},
  year={2025},
  publisher={IEEE}
}

@INPROCEEDINGS{bombino,
  author={Bombino, Andrea and Grimaldi, Simone and Mahmood, Aamir and Gidlund, Mikael},
  booktitle={IEEE WFCS}, 
  title={Machine Learning-Aided Classification Of {LoS/NLoS} Radio Links In Industrial {IoT}}, 
  year={2020},
  volume={},
  number={},
  pages={1-8},}

@article{olejniczak2024and,
  title={{LOS and NLOS} identification in real indoor environment using deep learning approach},
  author={Olejniczak, Alicja and Blaszkiewicz, Olga and Cwalina, Krzysztof K and Rajchowski, Piotr and Sadowski, Jaroslaw},
  journal={Digital Communications and Networks},
  volume={10},
  number={5},
  pages={1305--1312},
  year={2024},
  publisher={Elsevier}
}

@article{liu2022uwb,
  title={{UWB LOS/NLOS} identification in multiple indoor environments using deep learning methods},
  author={Liu, Qingzhi and Yin, Zhendong and Zhao, Yanlong and Wu, Zhilu and Wu, Mingyang},
  journal={Physical Communication},
  volume={52},
  pages={101695},
  year={2022},
  publisher={Elsevier}
}

@book{griva2008linear,
  title={Linear and nonlinear optimization 2nd edition},
  author={Griva, Igor and Nash, Stephen G and Sofer, Ariela},
  year={2008},
  publisher={SIAM}
}

@inproceedings{cristianini2001alignment,
  author    = {Nello Cristianini and John Shawe-Taylor and Andr{\'e} Elisseeff and Jaz S. Kandola},
  title     = {On Kernel-Target Alignment},
  booktitle = {Advances in Neural Information Processing Systems 14},
  editor    = {Thomas G. Dietterich and Suzanna Becker and Zoubin Ghahramani},
  volume    = {14},
  pages     = {367--373},
  publisher = {MIT Press},
  address   = {Cambridge, MA, USA},
  year      = {2001}
}

@article{cortes2012cka,
  author  = {Corinna Cortes and Mehryar Mohri and Afshin Rostamizadeh},
  title   = {Algorithms for Learning Kernels Based on Centered Alignment},
  journal = {Journal of Machine Learning Research},
  volume  = {13},
  number  = {28},
  pages   = {795--828},
  year    = {2012}
}

@book{HastieESL,
  author    = {T. Hastie and R. Tibshirani and J. Friedman},
  title     = {The Elements of Statistical Learning: Data Mining, Inference, and Prediction},
  edition   = {2},
  publisher = {Springer},
  year      = {2009}
}

@book{BishopPRML,
  author    = {C. M. Bishop},
  title     = {Pattern Recognition and Machine Learning},
  publisher = {Springer},
  year      = {2006}
}

@article{breiman2001random,
  title={Random forests},
  author={Breiman, Leo},
  journal={Machine learning},
  volume={45},
  number={1},
  pages={5--32},
  year={2001},
  publisher={Springer}
}

@article{abramson1982,
  author  = {I. S. Abramson},
  title   = {On Bandwidth Variation in Kernel Estimates---A Square Root Law},
  journal = {The Annals of Statistics},
  volume  = {10},
  number  = {4},
  pages   = {1217--1223},
  year    = {1982}
}

@article{Shrestha2025DLReceiverSurvey,
  author  = {Shree Krishna Shrestha and Syed M. Zafaruddin and H. Vincent Poor},
  title   = {Deep Learning in Wireless Communication Receiver: A Survey},
  journal = {arXiv preprint arXiv:2501.17184},
  year    = {2025},
  doi     = {10.48550/arXiv.2501.17184}
}

@Inbook{Perlich2017,
author="Perlich, Claudia",
editor="Sammut, Claude
and Webb, Geoffrey I.",
title="Learning Curves in Machine Learning",
bookTitle="Encyclopedia of Machine Learning and Data Mining",
year="2017",
publisher="Springer US",
address="Boston, MA",
pages="708--711",
}

@article{girolami2023experimental,
  title={An experimental evaluation based on direction finding specification for indoor localization and proximity detection},
  author={Girolami, Michele and Mavilia, Fabio and Furfari, Francesco and Barsocchi, Paolo},
  journal={IEEE J. Indoor Seamless Position. Navig.},
  volume={2},
  pages={36--50},
  year={2023},
  publisher={IEEE}
}

\end{document}